\documentclass[prb,twocolumn,superscriptaddress,floats]{revtex4}

\usepackage[dvips]{graphicx}
\usepackage{amsmath}
\usepackage{booktabs}
\usepackage{dcolumn}
\usepackage{sidecap}

\usepackage{rotating}

\newlength{\La} 

\newlength{\Lb} 

\newlength{\Lc} 

\newcolumntype{d}{D{.}{.}{-1}}

\newcommand{\Ang}{$\text{\AA}$}

\newcommand{\tn}{$T_\text{N}$}

\newcommand{\muB}{$\mu_B$}

\newcommand{\ruo}{RuO$_{6}$}

\newcommand{\sro}{Sr$_2$RuO$_4$}

\newcommand{\cro}{Ca$_2$RuO$_4$}

\newcommand{\csrx}{Ca$_{2-x}$Sr$_{x}$RuO$_4$}

\newcommand{\Csr}[1]{\ifthenelse{\equal{#1}{197}}{$\text{Ca}_\text{1.97}\text{Sr}_\text{0.03}\text{RuO}_\text{4}\,$}
{\ifthenelse{\equal{#1}{194}}{$\text{Ca}_\text{1.94}\text{Sr}_\text{0.06}\text{RuO}_\text{4}\,$}
{\ifthenelse{\equal{#1}{195}}{$\text{Ca}_\text{1.95}\text{Sr}_\text{0.05}\text{RuO}_\text{4}\,$}
{\ifthenelse{\equal{#1}{190}}{$\text{Ca}_\text{1.9}\text{Sr}_\text{0.1}\text{RuO}_\text{4}\,$}
{\ifthenelse{\equal{#1}{185}}{$\text{Ca}_\text{1.85}\text{Sr}_\text{0.15}\text{RuO}_\text{4}\,$}
{\ifthenelse{\equal{#1}{180}}{$\text{Ca}_\text{1.8}\text{Sr}_\text{0.2}\text{RuO}_\text{4}\,$}
{\ifthenelse{\equal{#1}{170}}{$\text{Ca}_\text{1.7}\text{Sr}_\text{0.3}\text{RuO}_\text{4}\,$}
{\ifthenelse{\equal{#1}{150}}{$\text{Ca}_\text{1.5}\text{Sr}_\text{0.5}\text{RuO}_\text{4}\,$}
{\ifthenelse{\equal{#1}{100}}{$\text{Ca}_\text{1.0}\text{Sr}_\text{1.0}\text{RuO}_\text{4}\,$}
{\ifthenelse{\equal{#1}{x}}{$\text{Ca}_\text{2-x}\text{Sr}_\text{x}\text{RuO}_\text{4}\,$}{XXXXX}}}}}}}}}}}

\begin{document}

\advance\vsize by 2 cm

\title{High pressure diffraction studies on \cro}

\author{P. Steffens }
\affiliation{II. Physikalisches Institut, Universit\"at zu K\"oln, Z\"ulpicher Str. 77, D-50937 K\"oln, Germany}

\author{O. Friedt}
\affiliation{II. Physikalisches Institut, Universit\"at zu K\"oln, Z\"ulpicher Str. 77, D-50937 K\"oln, Germany}
\affiliation{Laboratoire L\'eon Brillouin, C.E.A./C.N.R.S., F-91191 Gif-sur-Yvette CEDEX, France}

\author{P. Alireza}
\affiliation{Cavendish Laboratory, University of Cambridge, Madingley Road CB3 OHE, Cambridge, United Kingdom}

\author{W. G. Marshall}
\affiliation{ISIS Facility, Rutherford Appleton Laboratory, Chilton, Didcot, Oxon OX11 OQX, United Kingdom}

\author{W. Schmidt}
\affiliation{ILL, 6 Rue Jules Horowitz BP 156, 38042 Grenoble CEDEX 9, France}

\author{F. Nakamura}
\affiliation{Department of Quantum Matter, ADSM, Hiroshima University, Higashi-Hiroshima 739-8530, Japan}

\author{S. Nakatsuji}
\affiliation{Department of Physics, Kyoto University, Kyoto 606-8502, Japan}

\author{Y. Maeno}
\affiliation{Department of Physics, Kyoto University, Kyoto 606-8502, Japan}

\author{R. Lengsdorf}
\affiliation{II. Physikalisches Institut, Universit\"at zu K\"oln, Z\"ulpicher Str. 77, D-50937 K\"oln, Germany}

\author{M. M. Abd-Elmeguid}
\affiliation{II. Physikalisches Institut, Universit\"at zu K\"oln, Z\"ulpicher Str. 77, D-50937 K\"oln, Germany}

\author{M. Braden}
\affiliation{II. Physikalisches Institut, Universit\"at zu K\"oln, Z\"ulpicher Str. 77, D-50937 K\"oln, Germany}

\date{\today, \textbf{DRAFT}}

\pacs{PACS numbers:}

\begin{abstract}

We studied the crystal and magnetic structure of \cro\ by different diffraction techniques under high pressure. The
observed first order phase transition at moderate pressure (0.5~GPa) between the insulating phase and the metallic high
pressure phase is characterized by a broad region of phase coexistence. The following structural changes are observed
as function of pressure: a) a discontinuous change of both the tilt and rotation angle of the \ruo-Octahedra at this
transition, b) a gradual decrease of the tilt angle in the high pressure phase (p$>$0.5~GPa) and c) the disappearance
of the tilt above 5.5GPa leading to a higher symmetry structure. By single crystal neutron diffraction at low
temperature, the ferromagnetic component of the high pressure phase and a rearrangement of antiferromagnetic order in
the low pressure phase was observed.

\end{abstract}

\maketitle

\section{Introduction.}

Among the single-layered ruthenates, different topics have attracted research work during the recent years. A wide
variety of physical phenomena occur in these compounds that provide an interesting opportunity for the study of
competing ground states and the interplay betweeen structural, magnetic and transport properties. Starting from the
well-known unconventional superconductor \sro, which possesses the layered perovskite K$_2$NiF$_4$ structure, it is
only an isovalent substitution of Ca$^{2+}$ by Sr$^{2+}$ that leads to a variety of electronic and magnetic phases
\cite{nakatsuji,friedt01}and finally to \cro\, -- a Mott insulator with antiferromagnetic order (\tn=112~K) at low
temperature \cite{nakatsuji97,alexander,braden98,cao}.

\par One of the interesting phenomena that \cro\ displays is a
Mott transition to a metallic phase that occurs at T=357~K upon heating under ambient conditions. The insulating Mott
phase in \cro\ can also be suppressed by a moderate pressure \cite{nakamura}. At room temperature, \cro\ becomes
metallic at about 0.5~GPa. For slightly higher pressure it stays metallic down to lowest temperatures and exhibits
ferromagnetic order below T$\simeq$10~K. The Mott transition in both cases -- i.e. as function of temperature and of
pressure -- coincides with a structural transition. The electronic structure of \cro\, i.e. orbital occupation in its
insulating as well as in its metallic phase, is still under debate \cite{anisimov,fang,mizokawa,jung,lee,hotta}. In
view of the crossover from antiferro- to ferromagnetism the pressure dependence of \cro\ is of particular interest,
because it provides the possibility to study the interplay between the structure and magnetic degrees of freedom. It is
possible that the understanding of the ferromagnetic phase of \cro\ might also be helpful to explain the phenomena in
\Csr{x}, in which for 0.2$<$x$<$0.5 metamagnetic behaviour and an almost ferromagnetic state near $x$=0.5 is observed
\cite{nakatsuji03}. This article, describing the structure solution and direct observation of magnetic ordering by
neutron and x-ray diffraction, addresses the current lack of information concerning the nature of the high pressure
phase.

\section{Experimental.}

\begin{figure}[t]
\begin{center}
\includegraphics*[width=1.05\columnwidth]{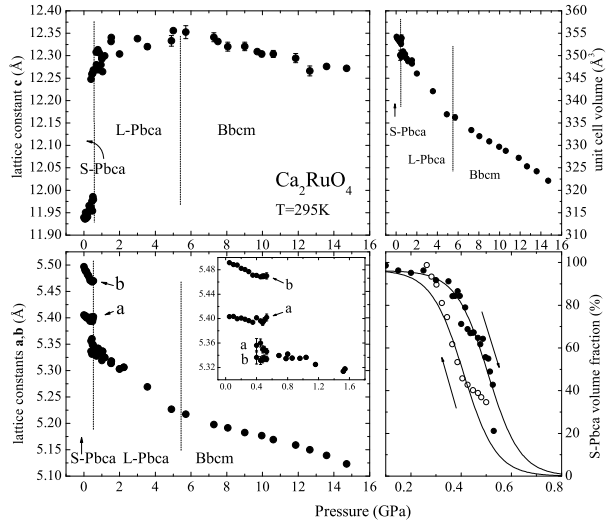}
\end{center}
\caption{Left: pressure dependence of the lattice constants of \cro. Data were taken by neutron powder diffraction (up
to 1~GPa only) and at HASYLAB, Hamburg (whole pressure range). The inset shows $a$ and $b$ near the phase transition
from \emph{S-Pbca} to \emph{L-Pbca}. Upper right: evolution of the unit cell volume. Lower right: hysteresis of the
\emph{S-Pbca} volume fraction at the transition. The corresponding space groups are indicated, the phase transitions
are marked by dashed lines.} \label{fig1}
\end{figure}

For the determination of the equation of state, powder samples of \cro\ were studied at the synchrotron beamline F3 at
HASYLAB in Hamburg and at the 3T.1 spectrometer at LLB, Saclay. At 3T.1 we used helium gas pressure cells and a
constant wavelength $\lambda=2.38$\Ang; the maximum pressure was 0.6~GPa. At F3 diamond anvil cells (Bridgman type,
0.6~mm diameter, pressure medium liquid nitrogen) were used and the lattice constants were calculated from the peak
positions in the energy dispersive powder diffraction spectra. The measurement covers the pressure region up to 15~GPa.
\par For the detailed structural analysis, Rietveld refinable data were
collected using the PEARL/HiPr time-of-flight diffractometer at ISIS on a powder sample up to a pressure of almost
10~GPa (Paris-Edinburgh-cell). As the pressure was not measured directly in this setup, the pressure was obtained by a
fit of the lattice constants to the equation of state (Fig.\,\ref{fig1}). All these measurements were done at room
temperature.
\par On two different single crystals, elastic neutron diffraction
was measured on the IN12 triple axis spectrometer at the ILL, Grenoble, with a clamp cell. These measurements were
performed in (100)-(010) and in (100)-(001) geometry (scattering plane), respectively . They covered the temperature
range between 1.5~K and room temperature. Crystal sizes in this case were only a few mm$^3$ -- note that up to now
large \cro\ single crystals are generally not available because they usually do not withstand the discontinuous
structural transition at 357~K. By the use of a triple axis instrument the relatively high background caused by the
pressure cell was significantly reduced.

\section{Results and Discussion.}

\subsection{Pressure dependence of the crystal structure at room temperature.\label{3a}}
\par As the first step of the structural analysis the pressure
dependence of the lattice constants, i.~e. the equation of state, was determined up to 15~GPa. The results are shown in
Fig.\,\ref{fig1}. In upstroke, one observes at 0.5~GPa the discontinuous transition from the \emph{S-Pbca} into the
\emph{L-Pbca} phase. Our structural analysis (see below) shows that it is well justified to call these phases (named
according to the short, respectively long lattice constant $c$) exactly like the phases which appear in the phase
diagram of \csrx\ \cite{braden98, friedt01}, because this transition is very similar to the one which occurs in \cro\
upon heating above 357K at ambient pressure. It is of strong first order character and displays a hysteresis of
$\sim$0.1~GPa (transition around 0.4~GPa in downstroke). At this transition, the lattice volume decreases by about
0.85\%, which is slightly more than at the temperature driven transition at ambient pressure \cite{friedt01}. So this
transition has to be regarded as just the same transition as the temperature-driven one which is susceptible to
pressure because of its negative volume change. Above this first-order transition, the orthorhombic splitting has
opposite sign ($a>b$), but the difference is very small and below the limit of detection in the synchrotron spectra;
however the PEARL/HiPr neutron diffraction results (see table) yield a splitting $\frac{a-b}{a+b}$ around $10^{-3}$.
Although it is obvious that pressure stabilizes the state with lower volume, the discontinuous increase of $c$ and the
positive $\frac{\partial c}{\partial p}$ over a wide pressure range are remarkable.

\begin{figure}[t]
\includegraphics*[width=0.77\columnwidth]{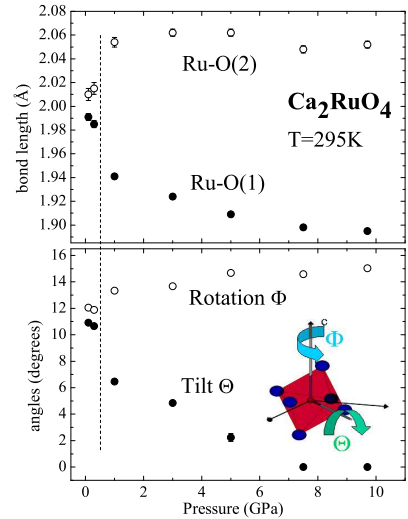}
\caption{Top: Evolution of the Ru-O bond lengths. Lower panel: Rotation and tilt angles of the \ruo-octahedra. The tilt
angle shown is the average of $mit{\Theta}$(O2) and $\mit{\Theta}$(O1). (Note that the pressure range is smaller than
in Figure \ref{fig1}.) \label{fig3}}
\end{figure}

\par The results of the full structure refinement of the data
collected at PEARL/HiPr are given in Table \ref{tab1}. For the refinement, the programs GSAS \cite{gsas} and Fullprof
\cite{fullprof} were used, and the results checked for consistency. During the refinement, special attention was paid
to the determination of the characteristic structural distortions, i.\,e. the rotation angle of the \ruo-octahedra
around their vertical axis ($\parallel$ c), $\mit{\Phi}$, and the tilt angle $\mit{\Theta}$ around an axis in the basal
plane ($\parallel$b and parallel to two of the O-O-bonds; for further details see ref. [\onlinecite{friedt01}]). The
results are summarized in Fig.\,\ref{fig3}.

\par In the high pressure region, where the tilt distortion
is only small, we described the spectra with different structural models, in which a tilt of the octahedra, caused by a
nonzero $z$-position of the O(1)-atom (basal oxygen) and nonzero $x$ and $y$ of O(2) (apical oxygen), was allowed or
forbidden, respectively. It turned out that above 7~GPa there was no significant difference in the R-values, which
would justify the lower symmetry. It was also possible to fix the horizontal Ca-positions to zero without affecting the
quality of the fit. In contrast, at pressures lower than 5~GPa satisfactory descriptions can only be obtained in the
usual \textit{Pbca} symmetry. At 5~GPa the difference between the two models is only small, but the calculated angles
are still significantly above zero.

\par
\begin{figure}[t]
\begin{center}
\includegraphics*[width=1.1\columnwidth]{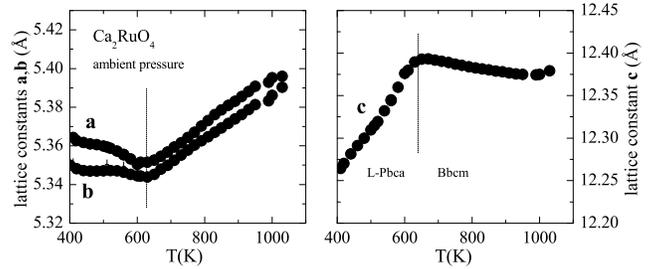}
\end{center}
\caption{Temperature dependence of the lattice constants of \cro\, at ambient pressure as measured by x-ray
diffraction.} \label{fig2}
\end{figure}

 With respect to Fig.\,\ref{fig1}, one may argue that the lattice
constants $a$, $b$ and $c$ and as well the cell volume display an anomaly between 5 and 6~GPa: a maximum in $c$ and a
kink in $a$, $b$ and $V$. Below, the compressibility is $-\frac{1}{V}\frac{\partial V}{\partial
p}=9.0\cdot10^{-3}\,\text{GPa}^{-1}$, and above, it is only $4.6\cdot10^{-3}\,\text{GPa}^{-1}$. This behavior is
qualitatively very similar to that observed upon heating at ambient pressure, see Fig.\,\ref{fig2}, where \cro\,
undergoes a second transition to a higher symmetric phase at 650~K\cite{steffens-unpub}. This analogy allows a
relatively precise determination of the pressure at which the phase transition occurs from the data in
Fig.\,\ref{fig1}. We therefore conclude that close to 5.5~GPa the tilt completely vanishes and \cro\ undergoes a
continuous phase transition to a phase with space group \textit{Bbcm} (standard setting \textit{Acam}), which is
characterized by a rotation of the \ruo-octahedra around the vertical axis only. This is the distortion pattern which
is also found in Gd$_2$CuO$_4$ \cite{gd2cuo4}. Note that the space group of this phase has to be distinguished from the
tetragonal space group \emph{I4$_1$/acd}, which is found \cite{friedt01} in \csrx\ and which is also characterized by
the \ruo\, rotation around $c$, because the sense of the rotation with respect to second nearest neighbor planes is
different. (This leads to a twice as long lattice constant $c$ in the case of \emph{I4$_1$/acd}, $c\simeq25\text{\AA}$,
compared to \emph{Bbcm} and \emph{Pbca}.) Another difference is that in the orthorhombic space group \emph{Bbcm} a very
small orthorhombicity seems to remain. Due to the absence of the tilt, this reflects the deformation of the octahedron
basal plane. The symmetry also allows for a splitting of the in-plane Ru-O bond lengths. Whether this is the case,
could not unambiguously be determined \cite{bem3}. In contrast to the tilt, the rotational distortion is quite
insensitive to pressure; it even increases slightly (Fig.\,\ref{fig3}).

\par The uncommon behaviour of the lattice constant $c$ is reflected in the bond length
between Ru and O(2), which increases under pressure up to 5~GPa, while the Ru-O(1) bond length
decreases continuously. This is obviously correlated to the observed pressure dependence of the
resistivity $\rho_{ab}$ and $\rho_c$ ($\rho_c$ increasing, $\rho_{ab}$ decreasing) \cite{nakamura}
-- the enhanced overlap of the in-plane orbitals corresponds to the usual effect of pressure and
may explain the higher conductivity.

\par It is tempting to interpret the pressure dependencies of $\mit{\Theta}$
and $\mit{\Phi}$ in view of theoretical calculations that have been carried out on the \Csr{x}\ phase diagram
\cite{anisimov,fang}. In their experiment, Nakamura \textit{et al.} \cite{nakamura} determined the evolution of the
ferromagnetic ordering temperature (partly deduced from resistivity measurements) with pressure. It increases from 10~K
just above the phase transition to its maximum of 25~K at a pressure $\sim$5~GPa, i.\,e. the increase coincides with
the reduction of the tilt angle, and the pressure where the tilt angle finally vanishes is near the maximum of
$T_\text{C}$. Fang and Terakura \cite{fang} performed calculations considering the influence of structural distortions
on the magnetic properties. They find that the octahedron rotation favors a ferromagnetic ground state due to the
smaller bandwidth of Ru $d_{xy}$, whereas the tilting has the opposite effect. This fits well to the fact that the
maximum ordering temperature is at the same pressure where we find the tilt disappearing (at 300K). On the other hand,
our structural analysis shows that below 5 GPa there are relatively large tilt angles which obviously do not prevent
(ferro-)magnetic order. Near $T_\text{C}$, i.e. at low temperature we even expect the tilt angle to be larger, although
there is no structure determination at low temperature in this pressure region yet. The occurrence of ferromagnetism in
the tilt distorted structure is also remarkable in view of the fact that in the phase diagram of \csrx\ it is the
octahedron's tilt setting in below a critical Sr-concentration x=0.5 which induces antiferromagnetic correlations and
drives the system very rapidly away from a ferromagnetic instability \cite{nakatsuji,friedt01}.

\subsection{Temperature dependence of the crystal structure under pressure.}

\begin{figure}[t]
\includegraphics*[width=0.9\columnwidth]{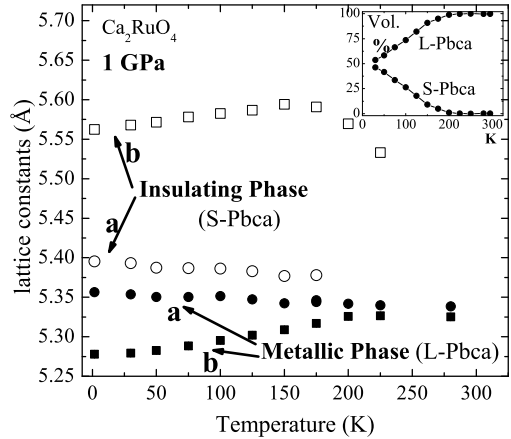}
\caption{Temperature dependence of the lattice constants and volume fractions (inset) of both phases at $P$=1~GPa. All
data were obtained on heating. In the mixed state, low temperature stabilizes the \textit{S-Pbca} phase and the phase
transition shifts to higher pressures: at 1~GPa and 1.5~K there is an equal mixture of both phases (inset). Therefore,
at low temperature both phases (\textit{L-} and \textit{S-Pbca}) could be studied simultaneously and at the same
pressure. \label{fig4}}
\end{figure}

\paragraph*{The two phase region.} The evolution of the lattice constants both of
the metallic and of the insulating phase as function of temperature was measured on a single crystal at a constant
pressure of approximately 1~GPa. At this pressure, there is a single metallic \emph{L-Pbca} phase at room temperature,
while at low temperatures the crystal partially transforms back into the insulating \emph{S-Pbca} phase (see inset of
Fig.\,\ref{fig4}), so there is a coexistence of both phases. As shown in Fig.\,\ref{fig1}, at room temperature the
hysteresis as function of pressure is $\sim$0.1~GPa, and a coexistence of both phases is found between 0.3 and 0.7~GPa.
At low temperatures this region is broader; in Raman studies by Snow \textit{et al.} \cite{snow}, the two phase state
could be observed up to $\sim$ 1~GPa, and measurements of magnetization and resistivity \cite{Nakamura-unpub} indicate
that small amounts of the insulating phase persist even up to $\sim$ 2~GPa. In this region of coexistence the lattice
constants of both phases could be measured as function of temperature at about 1~GPa (see Fig.\,\ref{fig4}).

\paragraph*{The insulating phase.}
In comparison with the low temperature lattice constants of \cro\ at ambient pressure,
$a=5.38\,$\Ang, $b=5.63\,$\Ang, the lattice is compressed only in $b$-direction, resulting in a
change of shape of the \ruo-octahedra; their basal planes are less elongated along $b$. As
mentioned above, Sr-doping has qualitatively the same effect. The origin of the maximum at 150~K
is not obvious. Near 150~K, however, is the N\'eel temperature of the pressurized
antiferromagnetic insulating phase (see below).

\paragraph*{The metallic phase.}
More interesting with respect to the discussion of (ferro-)magnetic order are the lattice constants of the metallic
phase. We find an unexpected splitting at low temperature \cite{bem2}. To determine which type of lattice distortion
occurs here, one needs further structural studies.

\subsection{Magnetism in the metallic (high pressure) phase.}

Remnant magnetizations  in the metallic high pressure phase of \cro\ are of the order of 0.4 $\mu_B$ per Ru (with
\textbf{\textit{M}} $\parallel$ \textbf{\textit{a}})\cite{nakamura,Nakamura-unpub}. We searched for scattering arising
from magnetic ordering in the metallic phase. Bragg scattering caused by ferromagnetic order is located right on the
nuclear Bragg reflections; this fact excluded a quantitative determination of the ferromagnetic ordered moment.
Nevertheless, the 004-reflection, which has a low nuclear structure factor and offers the best ratio of magnetic to
nuclear intensity, displayed a change when heated across the magnetic ordering temperature: with the statistics
achievable in our experiment, we obtained a decrease of 1.7 times the uncertainty $\sigma$ of the integrated intensity
between 1.5 and 20~K (at a pressure of $\sim$1.5~GPa). Within our experimental accuracy, this is consistent with the
value of the remnant magnetization.

\par We also searched for magnetic order at several other positions in reciprocal space. Among the related ruthenates like \csrx\
and Ti-doped \sro\, very different magnetic instabilities are found. Ti-doped \sro\ has static incommensurate magnetic
order \cite{braden02} at a wave vector (0.7,0.3,0); paramagnetic \Csr{150} has a very high susceptibility and displays
strongly enhanced magnetic fluctuations \cite{friedt04} at (0.22,0,0) as well as at the ferromagnetic zone center.

\par In contrast to macroscopic measurement methods, the direct observation by neutron scattering can give the information
whether there are other types of magnetic order than ferromagnetism which fully or partially determine the observed
magnetic properties. However, although measured at several positions in reciprocal space, neither type of
incommensurate or antiferromagnetic order could be detected. We estimate that long range order with an ordered moment
greater than $\sim$0.25 \muB\ would have been detectable.

\par The exclusion of other scenarios that might appear likely is an additional support that \cro\ really exhibits
ferromagnetic order. Other strong arguments \cite{nakamura} in favor of itinerant ferromagnetism are the size of the
macroscopic magnetic moment, which is inconsistent with a possible weak ferromagnetic component accompanying antiferro-
or incommensurate magnetic order. The hysteresis is observed in all lattice directions, and the ordering temperatures
are one order of magnitude lower than the N\'eel temperatures in \cro. Finally, the temperature dependence of the
resistivity indicates 2D itinerant ferromagnetism \cite{Nakamura-unpub,hatatani}.

\par Our examination of magnetic order by elastic neutron scattering supports
this scenario of itinerant ferromagnetism as the only one that can account for the magnetic properties of pressurized
\cro.

\subsection{Magnetism in the insulating (low pressure) phase.}

\par A further result concerns the \textit{S-Pbca} phase and its
antiferromagnetic order (\tn=112~K at $P$=0~GPa). The ordered moment on the Ru sites points along the $b$-direction of
the lattice \cite{bem1}. Early studies by neutron powder diffraction \cite{braden98} found a mixture of two different
antiferromagnetic phases. The alignment of the spin (\textbf{\textit{M}}=$1.3\mu_B$) is parallel to the $b$-direction
in both cases, but they differ in the stacking sequence of adjacent layers. This leads either to an A-centered or a
B-centered magnetic unit cell. The A-centered type of magnetic order ($T_\text{N}=112\textrm{K}$) was found to be the
majority phase at ambient pressure, while the minority B-phase had the higher \tn: $T_\text{N}=150\textrm{K}$. In some
other samples, a similar phase mixture was observed, but in all single crystals of high quality only a single
transition at 112~K could be detected \cite{Nakamura-unpub}. The appearance of the second phase is most likely caused
by small variations in oxygen content -- note that \cro\, with excess oxygen was shown to have \emph{only} the
B-centered antiferromagnetic order \cite{braden98}. Samples containing Sr (\csrx\ with $\geq0.03$) also have only the
B-centered order \cite{friedt01,steffens-unpub} and N\'eel temperatures of $\sim$150~K.

\par At 1~GPa and 1.5~K we were not able to detect any
magnetic scattering from the A-centered antiferromagnetic order in the \textit{S-Pbca} phase on any of the
corresponding Bragg-positions accessible in both crystal orientations. Instead, magnetic scattering was only detectable
in the 012-reflection, which originates from magnetic order of the B-centered type. We conclude that there is a single
magnetic phase of this type, in contrast to the behavior of the \textit{S-Pbca} phase at ambient pressure. We did not
determine the full temperature dependence of the magnetic scattering corresponding to the B-type antiferromagnetic
order, but at $T\simeq$150~K these reflections had entirely disappeared. Susceptibility measurements
\cite{Nakamura-unpub} show that \tn\ is 145~K at $P$=0.8~GPa, i.\,e. nearly the value which is found in all samples
with B-type order at ambient pressure.

\par The reason for this alteration of the magnetic order from A- to B-type is not obvious,
as the coupling between different layers is not yet well understood. However one may argue that this effect is the same
as that seen in the Sr-doped samples: very small amounts of Sr -- despite the bigger size of the Sr-ion -- have a
similar effect on the lattice constants and their temperature dependence as the application of pressure
\cite{steffens-unpub}, so the same subtle structural effect may be the reason for this.

\begin{figure}[t]
\includegraphics[scale=0.7]{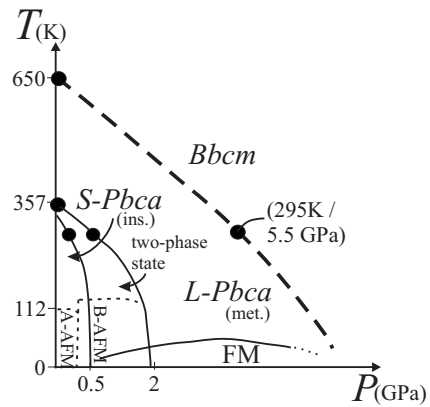}
\caption{Schematic structural and magnetic $P$-$T$ phase diagram of \cro\, (phase boundaries and distances not true to
scale). \emph{S-Pbca} and \emph{L-Pbca} denote the structure with tilt distortion (insulating and metallic
respectively) as described in the text, and at the transition a broad region of phase coexistence is found. The phase
border to the non-tilted Bbcm structure is known at two points only. A-AFM and B-AFM denote antiferromagnetic order
with A- and B-centered magnetic unit cell, respectively. FM is the ferromagnetic metallic phase [\onlinecite{nakamura}]
with maximum $T_\text{C}$ around 25~K at 5~GPa.} \label{fig5}
\end{figure}

\subsection{The phase diagram.}
In Fig. \ref{fig5} we show a schematic $P$-$T$ phase diagram of \cro\ which summarizes the results on structure and
magnetic order. At ambient pressure, \cro\ is an insulator with an A-centered antiferromagnetic unit cell below
\tn=112~K. By applying moderate pressure, the magnetic order is changed to B-centered antiferromagnetism. This change
occurs at some pressure between 0.2 and 0.8~GPa, and in the case of B-type order the N\'eel temperature is higher
(\tn$>$140~K). Upon increasing temperature or pressure, a first order structural phase transition takes place to the
\emph{L-Pbca} phase that is metallic and ferromagnetic at low temperature. This phase transition involves a relatively
broad region of coexistence of both phases -- at low temperature it extends from about 0.5 to 2~GPa; this region is
marked as "two-phase-state". At even higher temperatures or pressures, the tilt of the octahedra is completely
suppressed; the phase transition to the non-tilted structure with space group \emph{Bbcm} takes place at 650~K at
ambient pressure and at room temperature at 5.5~GPa.

\paragraph*{Comparison with the phase diagram of \csrx.} The application of pressure bears analogy to the effect of
a substitution of Ca by Sr. For 0$<$x$<$0.2, the Sr-concentration $x$ and the temperature control a first-order
transition from an insulating \emph{S-Pbca} phase, which is found at low $x$ and low temperature, to a metallic
\emph{L-Pbca} phase and finally via a second-order phase transition to \emph{Bbcm}
\cite{nakatsuji,friedt01,steffens-unpub}. Antiferromagnetism in the \emph{S-Pbca} phase changes from A- to B-centered
order upon Sr-doping. For 0$<$x$<$0.2, the $x$-$T$ phase diagram is therefore similar to the $P$-$T$ phase diagram for
pressures below $\sim$2~GPa, and the crystal structures are basically identical. However, although higher
Sr-concentrations finally suppress the tilt down to lowest temperatures, the further evolution is different, because in
the \csrx\ phase diagram the regions x$<$0.2 and x$>$0.2 are separated by a first-order phase boundary (due to
different rotation schemes, as discussed above in section \ref{3a}).

\section{Conclusion.}

\par We have determined the structural properties of
\cro\, up to a pressure of 10~GPa at room temperature and the lattice constants up to 15~GPa. The $P$-$T$ phase diagram
in Fig.\,\ref{fig5} summarizes the results. A first order structural transition leads from the insulating \emph{S-Pbca}
to the metallic \emph{L-Pbca} phase and involves a region in which both phases coexist. In the metallic phase, the tilt
angle is strongly reduced, but still significantly different from zero. Only at much higher pressure (above 5.5~GPa at
room temperature) the transition into the higher symmetry \emph{Bbcm}-structure without tilt was observed. In contrast,
the rotational distortion is only weakly pressure dependent. In the insulating phase two different types of
antiferromagnetic order exist, and pressure seems to control the crossover from A-centered at ambient pressure to
B-centered order (with higher N\'eel temperatures). In the metallic phase, we observed weak Bragg scattering due to
ferromagnetism. The unexpectedly large orthorhombic splitting towards low temperature in this phase indicates further
structural effects that have not yet been characterized in detail.

\section{Acknowledgemets.}

\par This work was supported by the Deutsche Forschungsgemeinschaft
through SFB 608 and grants from JSPS and MEXT of Japan.

\begin{table*}
\begin{tabular}{llllllll}
\hline \hline  $P$ (GPa)        &  0.1         &  0.3         &  1           &  3           &  5           &  7.5         &  9.7        \\
\hline Space group              & \emph{S-Pbca}& \emph{S-Pbca}& \emph{L-Pbca}& \emph{L-Pbca}& \emph{L-Pbca}& \emph{Bbcm}  & \emph{Bbcm} \\
\hline $a$ (\Ang)               &  5.4044(6)   &  5.4006(5)   &  5.3312(6)   &  5.2817(6)   &  5.2266(9)   &  5.2020(6)   &  5.1859(6)  \\
$b$ (\Ang)                      &  5.4904(6)   &  5.4760(5)   &  5.3160(6)   & 5.2689(5)    &  5.2187(9)   &  5.1865(6)   &  5.1673(6)  \\
$c$ (\Ang)                      &  11.9507(10) &  11.9664(9)  &  12.2923(8)  &  12.3354(5)  &  12.3541(6)  &  12.3301(6)  &  12.3078(6) \\
\hline $R_{wp}$ (\%)            &  3.33        &  3.07        &  2.70        &  2.25        &  2.72        &  2.40        &  2.39       \\
\hline \textbf{Ca} $x$          &  0.0108(13)  &  0.0092(13)  &  0.0116(13)  &  0.0109(14)  &  -0.001(2)   &  0           &  0          \\
$y$                             &  0.0456(12)  &  0.0463(12)  &  0.0289(15)  &  0.0155(19)  &  0.016(2)    &  0           &  0          \\
$z$                             &  0.3492(5)   &  0.3484(5)   &  0.3473(4)   &  0.3461(3)   &  0.3463(3)   &  0.3457(3)   &  0.3451(3)  \\
$U_{iso}$ (\Ang$^2$)            &  0.0102(15)  &  0.0145(17)  &  0.0107(11)  &  0.0118(10)  &  0.0065(11)  &  0.0083(8)   &  0.0093(8)  \\
\textbf{Ru} $U_{iso}$ (\Ang$^2$)&  0.0024(10)  &  0.0013(8)   &  0.0011(7)   &  0.0020(6)   &  0.0018(8)   &  0.0032(7)   &  0.0040(8)  \\
\textbf{O(1)} $x$               &  0.1953(6)   &  0.1966(6)   &  0.1909(4)   &  0.1894(3)   &  0.1847(4)   &  0.1851(4)   &  0.1832(4)  \\
$y$ (=0.5-$x$)                  &  0.3030(6)   &  0.3019(6)   &  0.3094(4)   &  0.3110(3)   &  0.3155(4)   &  0.3153(4)   &  0.3173(4)  \\
$z$                             &  0.0244(4)   &  0.0238(4)   &  0.0134(4)   &  0.0101(4)   &  0.0039(10)  &  0           &  0          \\
$U_{iso}$ (\Ang$^2$)            &  0.0082(10)  &  0.0080(9)   &  0.0077(8)   &  0.0070(7)   &  0.0058(7)   &  0.0071(6)   &  0.0070(6)  \\
\textbf{O(2)} $x$               &  -0.0602(7)  &  -0.0589(7)  &  -0.0385(8)  &  -0.0283(9)  &  -0.0163(17) &  0           &  0          \\
$y$                             &  -0.0201(10) &  -0.0186(10) &  -0.0116(12) &  -0.0103(13) &  -0.004(2)   &  0           &  0          \\
$z$                             &  0.1657(4)   &  0.1660(4)   &  0.1662(3)   &  0.1666(2)   &  0.1667(3)   &  0.1661(2)   &  0.1667(2)  \\
$U_{iso}$ (\Ang$^2$)            &  0.0081(11)  &  0.0098(11)  &  0.0092(10)  &  0.0091(8)   &  0.0084(8)   &  0.0095(6)   &  0.0095(6)  \\
\hline Ru-O(1) (\Ang)           &  1.991(3)    &  1.985(3)    &  1.941(2)    &  1.924(2)    &  1.909(2)    &  1.898(2)    &  1.895(2)   \\
Ru-O(2) (\Ang)                  &  2.010(5)    &  2.015(5)    &  2.054(4)    &  2.062(3)    &  2.062(3)    &  2.048(3)    &  2.052(3)   \\
O(1)-O(1) $\parallel a$ (\Ang)  &  2.825(5)    &  2.818(4)    &  2.759(3)    &  2.729(3)    &  2.703(3)    &  2.688(3)    &  2.685(3)   \\
O(1)-O(1) $\parallel b$ (\Ang)  &  2.808(5)    &  2.798(4)    &  2.732(3)    &  2.711(3)    &  2.697(3)    &  2.680(3)    &  2.675(3)   \\
\hline $\mit{\Theta}$(O1)       &  12.0(2)     &  11.7(2)     &  7.0(2)      &  5.3(2)      &  2.0(5)      &  -           &  -          \\
$\mit{\Theta}$(O2)              &  9.83(12)    &  9.55(11)    &  5.99(12)    &  4.4(2)      &  2.4(3)      &  -           &  -          \\
$\mit{\Phi}$                    &  12.15(9)    &  11.89(9)    &  13.33(7)    &  13.67(5)    &  14.67(6)    & 14.59(5)     & 15.02(5)\\ \hline
\end{tabular}
\caption{\label{tab1}Results of the room temperature structure refinements of \cro\ under pressure at the PEARL/HiPr
beamline (ISIS). Errors of the last digit are given in parentheses and represent only statistical errors, not
systematic errors which may for instance arise from correlations between parameters. The position of the
\textrm{Ru}-atom is always (0, 0, 0). Anisotropic thermal parameters could not be refined, therefore only the $U_{iso}$
are given.}
\end{table*}


\begin{thebibliography}{}




\bibitem{nakatsuji} S. Nakatsuji and Y. Maeno, Phys. Rev. Lett. \textbf{84}, 2666 (2000); \; S. Nakatsuji and Y. Maeno, Phys. Rev.
B \textbf{62}, 6458 (2000).

\bibitem{friedt01} O. Friedt, M. Braden, G. Andr\'e, P. Adelmann, S. Nakatsuji and Y. Maeno, Phys. Rev. B \textbf{63}, 174432
(2001).

\bibitem{nakatsuji97} S. Nakatsuji, S. Ikeda and Y. Maeno, J. Phys. Soc. Jpn. \textbf{66}, 1868 (1997).

\bibitem{alexander} C. S. Alexander, G. Cao, V. Dobrosavljevic, S. McCall, J. E. Crow, E. Lochner and R. P. Guertin,
Phys. Rev. B \textbf{60}, R8422 (1999).


\bibitem{braden98} M. Braden, G. Andr\'e, S. Nakatsuji and Y. Maeno, Phys. Rev. B \textbf{58}, 847 (1998).


\bibitem{cao} G. Cao, S. McCall, M. Shepard, J.E. Crow and R.P. Guertin, Phys. Rev. B \textbf{56}, R2916 (1997).


\bibitem{nakamura} F. Nakamura, T. Goko, M. Ito, T. Fujita, S. Nakatsuji, H. Fukazawa, Y. Maeno, P. Alireza, D. Forsythe and
S. R. Julian, Phys. Rev. B \textbf{65}, 220402 (2002).


\bibitem{anisimov} V. I. Anisimov, I. A. Nekrasov, D. E. Kondakov, T. M. Rice and M. Sigrist, Eur. Phys. J. B \textbf{25}, 191
(2002).


\bibitem{fang} Z. Fang and K. Terakura, Phys. Rev. B \textbf{64}, 020509(R) (2001); \; Z. Fang, N. Nagaosa and K. Terakura,
Phys. Rev. B \textbf{69}, 045116 (2004).


\bibitem{hotta} T. Hotta and E. Dagotto, Phys. Rev. Lett. \textbf{88}, 017201 (2002).

\bibitem{lee}  J. S. Lee, Y. S. Lee, T. W. Noh, S.-J. Oh, Jaejun Yu, S. Nakatsuji, H. Fukazawa, and Y. Maeno, Phys. Rev.
Lett. \textbf{89}, 257402 (2002).

\bibitem{jung} J. H. Jung, Z. Fang, J. P. He, Y. Kaneko, Y. Okimoto, and Y. Tokura, Phys. Rev. Lett. \textbf{91},
056403 (2003).

\bibitem{mizokawa} T. Mizokawa, L. H. Tjeng, G. A. Sawatzki, G. Ghiringhelli, O. Tjernberg, N. B. Brookes, H. Fukazawa, S. Nakatsuji
and Y. Maeno, Phys. Rev. Lett. \textbf{87}, 077202 (2001); \; T. Mizokawa, L. H. Tjeng, H.-J. Lin, C. T. Chen, S.
Schuppler, S. Nakatsuji, H. Fukazawa, and Y. Maeno, Phys. Rev. B \textbf{69}, 132410 (2004).


\bibitem{nakatsuji03} S. Nakatsuji, D. Hall, L. Balicas, Z. Fisk, K. Sugahara, M. Yoshioka and Y. Maeno, Phys. Rev. Lett. \textbf{90},
137202 (2003).

\bibitem{gsas} A.C. Larson and R.B. von Dreele, Los Alamos National Laboratory Report LAUR 86-748 (2000).

\bibitem{fullprof} J. Rodriguez-Carvajal, Physica B \textbf{192}, 55 (1993).

\bibitem{steffens-unpub} P. Steffens \textit{et al.}, to be published.

\bibitem{gd2cuo4} M. Braden, W. Paulus, A. Cousson, P. Vigoureux, G. Heger, A. Goukassov, P. Bourges and D. Petitgrand,
Europhys. Lett. \textbf{25}, 625 (1994).

\bibitem{snow} C. S. Snow, S. L. Cooper, G. Cao, J. E. Crow, H. Fukazawa, S. Nakatsuji and Y. Maeno, Phys. Rev. Lett.
\textbf{89}, 226401 (2002).

\bibitem{Nakamura-unpub} F. Nakamura \textit{et al.}, to be published.

\bibitem{braden02} M. Braden, O. Friedt, Y. Sidis, P. Bourges, M. Minataka and Y. Maeno, Phys. Rev. Lett. \textbf{88}, 197002
(2002); \; M. Minataka and Y. Maeno, Phys. Rev. B \textbf{63}, 180504 (2001).

\bibitem{friedt04} O. Friedt, P. Steffens, M. Braden, Y. Sidis, S. Nakatsuji and Y. Maeno, Phys. Rev. Lett. \textbf{93}, 147404
(2004).

\bibitem{hatatani} M. Hatatani and T. Moriya, J. Phys. Soc. Jpn. \textbf{64}, 3434 (1995).

\bibitem{bem3} An unconstrained shift of the O(1) $x$ and $y$ positions can result in two different Ru-O(1) bond lengths. This
is an important issue in the discussion of magnetism, because the distortion pattern arising from this could stabilize
antiferroorbital order. Unconstrained refinement yields sizeable splittings up to two percent, but the quality of the
fit is \emph{not} affected by enforcing the O(1)-shift to be perpendicular to the bond, i.\,e. a single Ru-O(1) bond
length. The table therefore lists the results of the refinement with equal Ru-O(1) bonds. In the case of the S-Pbca
structure at ambient pressure the same question arises -- it could be solved by high resolution powder and single
crystal studies which did not find significant splitting, see [\onlinecite{braden98,friedt01}].

\bibitem{bem1} Note that due to the rotation of the octahedra, the Ru magnetic moment is also slightly canted away from the
crystallographic b-axis.

\bibitem{bem2} Here, one has to take account of the possibility, that due to the phase mixture uncontrollable
real structure effects influence the results if the domain sizes are very small. However, we consider this as unlikely,
as the analysis of the peak widths did not yield any indication for this. In addition, a similar splitting was observed
on different samples.



\end{thebibliography}
\end{document}